
\newif\iflanl
\openin 1 lanlmac
\ifeof 1 \lanlfalse \else \lanltrue \fi
\closein 1
\iflanl
    \input lanlmac
\else
    \message{[lanlmac not found - use harvmac instead}
    \input harvmac
\fi
\newif\ifhypertex
\ifx\hyperdef\UnDeFiNeD
    \hypertexfalse
    \message{[HYPERTEX MODE OFF}
    \def\hyperref#1#2#3#4{#4}
    \def\hyperdef#1#2#3#4{#4}
    \def\hypernoname{}
    \def\e@tf@ur#1{}
    \def\hep-th/#1#2#3#4#5#6#7{{\tt hep-th/#1#2#3#4#5#6#7}}
    \def\CERN{\address{CERN, Geneva, Switzerland}}
\else
    \hypertextrue
    \message{[HYPERTEX MODE ON}
  \def\hep-th/#1#2#3#4#5#6#7{
  {\tt hep-th/#1#2#3#4#5#6#7}}
\def\CERN{\address{
Theory Division, CERN, Geneva, Switzerland}}
\fi
\newif\ifdraft

\noblackbox
\catcode`\@=11
\newif\iffrontpage
\ifx\answ\bigans
\def\titleft{\titsm}
\magnification=1200\baselineskip=14pt plus 2pt minus 1pt
%
\advance\hoffset by-0.075truein
\hsize=6.15truein\vsize=600.truept\hsbody=\hsize\hstitle=\hsize
\else\let\lr=L
\def\titleft{\titla}
\magnification=1000\baselineskip=14pt plus 2pt minus 1pt
%
\hoffset=-0.75truein\voffset=-.0truein
\vsize=6.5truein
\hstitle=8.truein\hsbody=4.75truein
\fullhsize=10truein\hsize=\hsbody
\fi
\parskip=4pt plus 15pt minus 1pt
%
\newif\iffigureexists
\newif\ifepsfloaded
\def\epsfcheck{
\ifdraft
\input epsf\epsfloadedtrue
\else
  \openin 1 epsf
  \ifeof 1 \epsfloadedfalse \else \epsfloadedtrue \fi
  \closein 1
  \ifepsfloaded
    \input epsf
  \else
\immediate\write20{NO EPSF FILE --- FIGURES WILL BE IGNORED}
  \fi
\fi
\def\epsfcheck{}}
\def\checkex#1{
\ifdraft
\figureexistsfalse\immediate%
\write20{Draftmode: figure #1 not included}
\else\relax
    \ifepsfloaded \openin 1 #1
	\ifeof 1
           \figureexistsfalse
  \immediate\write20{FIGURE FILE #1 NOT FOUND}
	\else \figureexiststrue
	\fi \closein 1
    \else \figureexistsfalse
    \fi
\fi}
\def\missbox#1#2{$\vcenter{\hrule
\hbox{\vrule height#1\kern1.truein
\raise.5truein\hbox{#2} \kern1.truein \vrule} \hrule}$}
\def\lfig#1{
\let\labelflag=#1%
\def\numb@rone{#1}%
\ifx\labelflag\UnDeFiNeD%
{\xdef#1{\the\figno}%
\writedef{#1\leftbracket{\the\figno}}%
\global\advance\figno by1%
}\fi{\hyperref{}{figure}{{\numb@rone}}{Fig.{\numb@rone}}}}
\def\figinsert#1#2#3#4{
\epsfcheck\checkex{#4}%
\def\figsize{#3}%
\let\flag=#1\ifx\flag\UnDeFiNeD
{\xdef#1{\the\figno}%
\writedef{#1\leftbracket{\the\figno}}%
\global\advance\figno by1%
}\fi
\goodbreak\midinsert%
\iffigureexists
\centerline{\epsfysize\figsize\epsfbox{#4}}%
\else%
\vskip.05truein
  \ifepsfloaded
  \ifdraft
  \centerline{\missbox\figsize{Draftmode: #4 not included}}%
  \else
  \centerline{\missbox\figsize{#4 not found}}
  \fi
  \else
  \centerline{\missbox\figsize{epsf.tex not found}}
  \fi
\vskip.05truein
\fi%
{\smallskip%
\leftskip 4pc \rightskip 4pc%
\noindent\ninepoint\sl \baselineskip=11pt%
{\bf{\hyperdef\hypernoname{figure}{{#1}}{Fig.{#1}}}:~}#2%
\smallskip}\bigskip\endinsert%
}
%
\font\bigit=cmti10 scaled \magstep1

\font\titla=cmr10 scaled\magstep3
\font\tenmss=cmss10
\font\absmss=cmss10 scaled\magstep1

\newfam\mssfam
\font\footrm=cmr8  \font\footrms=cmr5
\font\footrmss=cmr5   \font\footi=cmmi8
\font\footis=cmmi5   \font\footiss=cmmi5
\font\footsy=cmsy8   \font\footsys=cmsy5
\font\footsyss=cmsy5   \font\footbf=cmbx8
\font\footmss=cmss8
\def\footfont{\def\rm{\fam0\footrm}
\textfont0=\footrm \scriptfont0=\footrms
\scriptscriptfont0=\footrmss
\textfont1=\footi \scriptfont1=\footis
\scriptscriptfont1=\footiss
\textfont2=\footsy \scriptfont2=\footsys
\scriptscriptfont2=\footsyss
\textfont\itfam=\footi \def\it{\fam\itfam\footi}
\textfont\mssfam=\footmss \def\mss{\fam\mssfam\footmss}
\textfont\bffam=\footbf \def\bf{\fam\bffam\footbf} \rm}
\def\tenpoint{\def\rm{\fam0\tenrm}
\textfont0=\tenrm \scriptfont0=\sevenrm
\scriptscriptfont0=\fiverm
\textfont1=\teni  \scriptfont1=\seveni
\scriptscriptfont1=\fivei
\textfont2=\tensy \scriptfont2=\sevensy
\scriptscriptfont2=\fivesy
\textfont\itfam=\tenit \def\it{\fam\itfam\tenit}
\textfont\mssfam=\tenmss \def\mss{\fam\mssfam\tenmss}
\textfont\bffam=\tenbf \def\bf{\fam\bffam\tenbf} \rm}
\ifx\answ\bigans\def\abstractfont{\tenpoint}\else
\def\abstractfont{\def\rm{\fam0\absrm}
\textfont0=\absrm \scriptfont0=\absrms
\scriptscriptfont0=\absrmss
\textfont1=\absi \scriptfont1=\absis
\scriptscriptfont1=\absiss
\textfont2=\abssy \scriptfont2=\abssys
\scriptscriptfont2=\abssyss
\textfont\itfam=\bigit \def\it{\fam\itfam\bigit}
\textfont\mssfam=\absmss \def\mss{\fam\mssfam\absmss}
\textfont\bffam=\absbf \def\bf{\fam\bffam\absbf}\rm}\fi
%
\def\f@@t{\baselineskip10pt\lineskip0pt\lineskiplimit0pt
\bgroup\aftergroup\@foot\let\next}
\setbox\strutbox=\hbox{\vrule height 8.pt depth 3.5pt width\z@}
\def\vfootnote#1{\insert\footins\bgroup
\baselineskip10pt\footfont
\interlinepenalty=\interfootnotelinepenalty
\floatingpenalty=20000
\splittopskip=\ht\strutbox \boxmaxdepth=\dp\strutbox
\leftskip=24pt \rightskip=\z@skip
\parindent=12pt \parfillskip=0pt plus 1fil
\spaceskip=\z@skip \xspaceskip=\z@skip
\Textindent{$#1$}\footstrut\futurelet\next\fo@t}
\def\Textindent#1{\noindent\llap{#1\enspace}\ignorespaces}
\def\foot{\global\advance\ftno by1%
\attach{\hyperref{}{footnote}{\the\ftno}{\footsymbolgen}}%
\vfootnote{\hyperdef\hypernoname{footnote}{\the\ftno}{\footsymbol}}}%
\def\footnote#1{\global\advance\ftno by1%
\attach{\hyperref{}{footnote}{\the\ftno}{#1}}%
\vfootnote{\hyperdef\hypernoname{footnote}{\the\ftno}{#1}}}%
\newcount\lastf@@t           \lastf@@t=-1
\newcount\footsymbolcount    \footsymbolcount=0
\global\newcount\ftno \global\ftno=0
\def\footsymbolgen{\relax\footsym
\global\lastf@@t=\pageno\footsymbol}
\def\footsym{\ifnum\footsymbolcount<0
\global\footsymbolcount=0\fi
{\iffrontpage \else \advance\lastf@@t by 1 \fi
\ifnum\lastf@@t<\pageno \global\footsymbolcount=0
\else \global\advance\footsymbolcount by 1 \fi }
\ifcase\footsymbolcount
\fd@f\dagger\or \fd@f\diamond\or \fd@f\ddagger\or
\fd@f\natural\or \fd@f\ast\or \fd@f\bullet\or
\fd@f\star\or \fd@f\nabla\else \fd@f\dagger
\global\footsymbolcount=0 \fi }
\def\fd@f#1{\xdef\footsymbol{#1}}
\def\space@ver#1{\let\@sf=\empty \ifmmode #1\else \ifhmode
\edef\@sf{\spacefactor=\the\spacefactor}
\unskip${}#1$\relax\fi\fi}
\def\attach#1{\space@ver{\strut^{\mkern 2mu #1}}\@sf}
%
\newif\ifnref
\def\rrr#1#2{\relax\ifnref\nref#1{#2}\else\ref#1{#2}\fi}
\def\ldf#1#2{\begingroup\obeylines
\gdef#1{\rrr{#1}{#2}}\endgroup\unskip}
\def\nrf#1{\nreftrue{#1}\nreffalse}
\def\doubref#1#2{\refs{{#1},{#2}}}

\nreffalse
\def\refout{\listrefs}
%
\def\eqn#1{\xdef #1{(\noexpand\hyperref{}%
{equation}{\secsym\the\meqno}%
{\secsym\the\meqno})}\eqno(\hyperdef\hypernoname{equation}%
{\secsym\the\meqno}{\secsym\the\meqno})\eqlabeL#1%
\writedef{#1\leftbracket#1}\global\advance\meqno by1}
\def\eqnalign#1{\xdef #1{\noexpand\hyperref{}{equation}%
{\secsym\the\meqno}{(\secsym\the\meqno)}}%
\writedef{#1\leftbracket#1}%
\hyperdef\hypernoname{equation}%
{\secsym\the\meqno}{\e@tf@ur#1}\eqlabeL{#1}%
\global\advance\meqno by1}
\def\eqnalign#1{\xdef #1{(\secsym\the\meqno)}
\writedef{#1\leftbracket#1}%
\global\advance\meqno by1 #1\eqlabeL{#1}}
%
\def\hsect#1{\hyperref{}{section}{#1}{section~#1}}

\def\chap#1{\newsec{#1}}
\def\chapter#1{\chap{#1}}
\def\sect#1{\subsec{#1}}
\def\section#1{\sect{#1}}
\def\\{\ifnum\lastpenalty=-10000\relax
\else\hfil\penalty-10000\fi\ignorespaces}
\def\note#1{\leavevmode%
\edef\@@marginsf{\spacefactor=\the\spacefactor\relax}%
\ifdraft\strut\vadjust{%
\hbox to0pt{\hskip\hsize%
\ifx\answ\bigans\hskip.1in\else\hskip .1in\fi%
\vbox to0pt{\vskip-\dp
\strutbox\sevenbf\baselineskip=8pt plus 1pt minus 1pt%
\ifx\answ\bigans\hsize=.7in\else\hsize=.35in\fi%
\tolerance=5000 \hbadness=5000%
\leftskip=0pt \rightskip=0pt \everypar={}%
\raggedright\parskip=0pt \parindent=0pt%
\vskip-\ht\strutbox\noindent\strut#1\par%
\vss}\hss}}\fi\@@marginsf\kern-.01cm}
\def\titlepage{%
\frontpagetrue\nopagenumbers\abstractfont%
\hsize=\hstitle\rightline{\vbox{\baselineskip=10pt%
{\abstractfont\pubnum}}}\pageno=0}
\frontpagefalse
\def\pubnum{}
\def\pdate{\number\month/\number\yearltd}
\def\makefootline{\iffrontpage\vskip .27truein
\line{\the\footline}
\vskip -.1truein\leftline{\vbox{\baselineskip=10pt%
{\abstractfont\pdate}}}
\else\vskip.5cm\line{\hss \tenrm $-$ \folio\ $-$ \hss}\fi}
\def\title#1{\vskip .7truecm\titlestyle{\titleft #1}}
\def\titlestyle#1{\par\begingroup \interlinepenalty=9999
\leftskip=0.02\hsize plus 0.23\hsize minus 0.02\hsize
\rightskip=\leftskip \parfillskip=0pt
\hyphenpenalty=9000 \exhyphenpenalty=9000
\tolerance=9999 \pretolerance=9000
\spaceskip=0.333em \xspaceskip=0.5em
\noindent #1\par\endgroup }
\def\autskip{\ifx\answ\bigans\vskip.5truecm\else\vskip.1cm\fi}
\def\author#1{\vskip .7in \centerline{#1}}

\def\address#1{\ifx\answ\bigans\vskip.2truecm
\else\vskip.1cm\fi{\it \centerline{#1}}}
\def\abstract#1{
\vskip .5in\vfil\centerline
{\bf Abstract}\penalty1000
{{\smallskip\ifx\answ\bigans\leftskip 2pc \rightskip 2pc
\else\leftskip 5pc \rightskip 5pc\fi
\noindent\abstractfont \baselineskip=12pt
{#1} \smallskip}}
\penalty-1000}
\def\endpage{\tenpoint\supereject\global\hsize=\hsbody%
\frontpagefalse\footline={\hss\tenrm\folio\hss}}
%

\def\bfone{\relax{\rm 1\kern-.35em 1}}
\def\inbar{\vrule height1.5ex width.4pt depth0pt}
\def\IC{\relax\,\hbox{$\inbar\kern-.3em{\mss C}$}}
\def\ID{\relax{\rm I\kern-.18em D}}
\def\IF{\relax{\rm I\kern-.18em F}}
\def\IH{\relax{\rm I\kern-.18em H}}
\def\II{\relax{\rm I\kern-.17em I}}
\def\IN{\relax{\rm I\kern-.18em N}}
\def\IP{\relax{\rm I\kern-.18em P}}
\def\IQ{\relax\,\hbox{$\inbar\kern-.3em{\rm Q}$}}
\def\IR{\relax{\rm I\kern-.18em R}}
\font\cmss=cmss10 \font\cmsss=cmss10 at 7pt
\def\ZZ{\relax\ifmmode\mathchoice
{\hbox{\cmss Z\kern-.4em Z}}{\hbox{\cmss Z\kern-.4em Z}}
{\lower.9pt\hbox{\cmsss Z\kern-.4em Z}}
{\lower1.2pt\hbox{\cmsss Z\kern-.4em Z}}\else{\cmss Z\kern-.4em
Z}\fi}
\def\a{\alpha} \def\b{\beta} \def\d{\delta}
 
 \def\l{\lambda}
\def\L{\Lambda}

\def\nup#1({Nucl.\ Phys.\ $\us {B#1}$\ (}
\def\plt#1({Phys.\ Lett.\ $\us  {#1}$\ (}
\def\cmp#1({Comm.\ Math.\ Phys.\ $\us  {#1}$\ (}
\def\prp#1({Phys.\ Rep.\ $\us  {#1}$\ (}
\def\prl#1({Phys.\ Rev.\ Lett.\ $\us  {#1}$\ (}
\def\prv#1({Phys.\ Rev.\ $\us  {#1}$\ (}
\def\mpl#1({Mod.\ Phys.\ Let.\ $\us  {A#1}$\ (}
\def\ijmp#1({Int.\ J.\ Mod.\ Phys.\ $\us{A#1}$\ (}
\def\tit#1|{{\it #1},\ }
%

%

\def\ni{\noindent}
\def\tilde{\widetilde}
\def\bar{\overline}
\def\us#1{\underline{#1}}

\def\hat{\widehat}

\def\Coe#1.#2.{{#1\over #2}}
\def\coeff#1#2{\relax{\textstyle {#1 \over #2}}\displaystyle}
\def\coe#1.#2.{\relax{\textstyle {#1 \over #2}}\displaystyle}

\def\shalf{\relax{\textstyle {1 \over 2}}\displaystyle}

\def\notin{\hbox{{$\in$}\kern-.51em\hbox{/}}}

\def\ket#1{\,\big|\,#1\,\big>\,}

\def\del{\partial}

\def\nex#1{$N\!=\!#1$}

 \def\ie{{\it i.e.}}
\catcode`\@=12
\def\sc{superconformal\ }
\def\sca{\sc algebra\ }
\def\LG{Lan\-dau-Ginz\-burg\ }
\def\nul#1,{{\it #1},}

\def\ga#1{\gamma_{0,#1}}
\def\Ga#1{\Gamma_{0,#1}}

\def\L{\Lambda}

\def\brs{$BRST$}

\def\zw#1#2.{{#2\over(z-w)^{#1}}}

\def\ket#1{\,\big\vert#1\big\rangle}
\def\ket#1{\,\left\vert#1\right\rangle}
\def\cont{\coeff1{2\pi i}\!\oint dz\,}

\def\p#1{\psi_{#1}}\def\bp#1{\bar\psi_{#1}}
\def\df#1{\del\phi_{#1}}\def\bdf#1{\del\bar\phi_{#1}}
\def\vp{\varphi}
\def\g{\gamma}
\def\mod{{\rm mod}\,}
\def\M{n}
\ldf\TOPALG{E.\ Witten, \cmp{117} (1988) 353; \cmp{118} (1988) 411;
\nup340 (1990) 281.}
\ldf\EYtop{T.\ Eguchi and S.\ Yang, \mpl4 (1990) 1693.}
\ldf\LS{W.\ Lerche and A.\ Sevrin, {\it On the \LG Realization
of Topological Gravities}, preprint CERN-TH.7210/94,
\hep-th/9403183.}
\ldf\topgr{E.\ Witten,  \nup340 (1990) 281; E.\ and H.\ Verlinde,
\nup348 (1991) 457.}
\ldf\Witgr{E.\ Witten, \nup373 (1992) 187. }
\ldf\DVV{R.\ Dijkgraaf, E. Verlinde and H. Verlinde, \nup{352} (1991)
59.}
\ldf\Loss{A. Lossev, \nul{Descendants constructed from matter field
and K. Saito higher residue pairing in Landau-Ginzburg theories
coupled to topological gravity}, preprint TPI-MINN-92-40-T,
\hep-th/9211090.}
\ldf\BGS{B.\ Gato-Rivera and A.M.\ Semikhatov, \plt B293 (1992) 72,
\hep-th/9207004.}
\ldf\LVW{W.\ Lerche, C.\ Vafa and N.P.\ Warner, \nup324 (1989) 427.}
\ldf\Wred{K.\ Ito, \plt B259(1991) 73; \nup370(1992) 123,
\hep-th/9210143; D.\ Nemeschansky and S.\ Yankielowicz, {\it N=2
W-algebras, Kazama-Suzuki models and Drinfeld-Sokolov reduction},
preprint USC-91-005A; K.\ Ito, J.\ Madsen and J.\ Petersen,
\plt318(1993) 315, \hep-th/9207009; K.\ Ito and H.\ Kanno, {\it Lie
Superalgebra and Extended Topological Conformal Symmetry in
Non-critical $W_{3}$ Strings}, preprint UTHEP-277,
\hep-th/9405049.}
\ldf\DeNiA{D.\ Nemeschansky and N.P. \ Warner, {\it Refining the
Elliptic Genus},  USC preprint USC-94/002,
\hep-th/9403047.}
\ldf\DeNiB{D.\ Nemeschansky and N.P.\ Warner, in preparation.}
\ldf\topw{K.\ Li, \plt B251 (1990) 54, \nup346 (1990) 329; H.\ Lu,
C.N.\ Pope and X.\ Shen, \nup366(1991) 95; S.\ Hosono, \nul{
Algebraic definition of topological W-gravity}, preprint UT-588;
H.\ Kunitomo, Prog.\ Theor.\ Phys.\ 86 (1991) 745.}
\ldf\KS{Y.\ Kazama and H.\ Suzuki, \nup321(1989) 232.}
\ldf\Keke{K.\ Li, \nup354(1991) 711; \nup354(1991)725.}
\ldf\Vafa{C.\ Vafa, \mpl6 (1991) 337.}
\ldf\dis{J.\ Distler, \nup342(1990) 523.}
\ldf\BLNWB{M.\ Bershadsky, W.\ Lerche, D.\ Nemeschansky and N.P.\
Warner, \nup401 (1993) 304, \hep-th/9211040.}
\ldf\EYQ{T.\ Eguchi, H.\ Kanno, Y.\ Yamada and S.-K.\ Yang, \plt B305
(1993) 235, \hep-th/9302048.}
\ldf\wground{P.\ Bouwknegt, J.\ McCarthy and K.\ Pilch, \nul{
Semi-infinite cohomology of $W$-algebras}, preprint USC-93/11 and
ADP-23-200/M15, \hep-th/9302086; \nul{On the W gravity
spectrum and its G structure}, preprint USC-93/27,
\hep-th/9311137; E.\ Bergshoeff, J.\ de Boer,
M.\ de Roo and T.\ Tjin, \nup420(1994)379, \hep-th/9312185.}
\ldf\twistKS{W.\ Lerche, \plt252B (1990) 349; T.\ Eguchi, S.\ Hosono
and S.K.\ Yang, \cmp140(1991) 159; T.\ Eguchi, T.\ Kawai, S.\
Mizoguchi and S.K.\ Yang, Rev.\ Math.\ Phys.\ 4 (1992) 329.}
\ldf\wbrs{M.\ Bershadsky, W.\ Lerche, D.\ Nemeschansky and N.P.\
Warner, \plt B292 (1992) 35, \hep-th/9207067; E.\
Bergshoeff, A.\ Sevrin and X.\ Shen, \plt B296 (1992) 95,
\hep-th/9209037; J. de Boer and J. Goeree, \nup405 (1993) 669,
\hep-th/9211108.}
\ldf\BLLS{A.\ Boresch, K.\ Landsteiner, W.\ Lerche and A.\ Sevrin,
{\it Topological strings from Hamiltonian reduction}, to appear.}
\ldf\LGfree{P.\ Fr\'e, L.\ Girardello, A.\ Lerda and P.\ Soriani,
\nup387 (1992) 333, \hep-th/9204041; E.\ Witten, {\it On the
Landau-Ginzburg description of N=2 minimal models}, preprint
IASSNS-HEP-93-1, \hep-th/9304026.}
\ldf\FMS{D.\ Friedan, E.\ Martinec and S.\ Shenker, \nup271(1986)
93.}
\ldf\BMP{P.\ Bouwknegt, J.\ McCarthy and K.\ Pilch,
{\it On physical states in 2d (topological) gravity},
preprint CERN-TH.6645/92.}
\ldf\topw{K.\ Li, \plt B251 (1990) 54, \nup346 (1990) 329, {\it
Linear $W_N$-gravity}, preprint CALT-68-1724; H.\ Lu, C.N.\ Pope and
X.\ Shen, \nup366(1991) 95; S.\ Hosono, \nul{ Algebraic definition of
topological W-gravity}, preprint UT-588; H.\ Kunitomo, Prog.\ Theor.\
Phys.\ 86 (1991) 745.}
\def\pubnum{
\hbox{CERN-TH.7442/94}
\hbox{USC-94/014}
\hbox{hep-th/9409069}}
\def\pdate{}
\titlepage
\title
{On the Algebraic Structure of Gravitational
Descendants in CP(n--1) Coset Models}
\vskip-1.1cm\autskip
\author{W.\ Lerche}
\CERN
\vskip-.7truecm
\author{N.P.\ Warner}
\address{Physics Department, U.S.C., University Park,
Los Angeles, CA 90089}
\vskip-1.truecm
\abstract{
We investigate how specific free-field realizations of twisted \nex2
supersymmetric coset models give rise to natural extensions of the
``matter'' Hilbert spaces in such a manner as to incorporate the
various gravitational excitations. In particular, we show that
adopting a particular screening prescription is equivalent to
imposing the requisite equivariance condition on cohomology. We find
a simple algebraic characterization of the $W_n$-gravitational ground
ring spectra of these theories in terms of affine-$SU(\M)$ highest
weights.}
\vfil
\vskip 1.cm
\ni CERN-TH.7442/94\hfill\break
\ni September 1994
\endpage
\baselineskip=14pt plus 2pt minus 1pt
\sequentialequations
\chapter{Introduction}
{}From the work of \doubref\BLNWB\LS\ it is evident that there should
be some beautiful and simple algebraic structure underlying the
coupling of topological $W$-matter theories to topological
$W$-gravity \topw. Our purpose in this paper is to simplify and then
generalize the results of \LS\ and thereby elucidate the algebraic
structure.

The key idea accomplishing this is to find a matter representation of
the gravitational descendants. Specifically, this means expanding the
unitary matter theory in a natural manner to a (necessarily)
non-unitary theory, and then finding representations of the
gravitational descendants within the extended matter Hilbert space.
The basic idea was first suggested in \Loss\ for topological
\LG theories coupled to ordinary topological gravity (it also
surfaced in the work of \BLNWB). This idea was then developed and
greatly refined in \EYQ, where the authors introduced a
computationally (and conceptually) invaluable simplification in the
form of a global Hilbert space rotation that maps the \brs\ operator
into a much simpler expression.

The problem in applying these techniques to more general models is to
find a systematic and natural extension of the matter Hilbert space
so as to incorporate the various kinds of gravitational
descendants. For our problem, where we consider topological
$W$-minimal models \doubref\KS\Wred\ coupled to $W$-gravity,
the natural Hilbert space extension is obtained by removing the
fermionic screening prescription in the Drinfeld-Sokolov construction
of these models. This particular prescription was suggested in
\BLNWB, which considered a specific Drinfeld-Sokolov reduction that
yields a non-standard free-field realization of these models. Though
this specific free-field realization was natural in the context
\wbrs\ of the problems discussed in \BLNWB, it was prohibitively
inconvenient for discussing the full ground ring \wground\ spectrum.
In the \LG formulation of \LS, on the other hand, the full spectrum
is relatively easy to obtain, but its algebraic structure remained
obscure.

In the present paper we will study the modified fermionic screening
prescription in the standard Drinfeld-Sokolov type free-field
realization, and show that it leads to a very simple description of
the $W_\M$-gravitational descendants directly in terms of weight
lattices of $SU(\M)$. Specifically, in the next section we will
briefly review the results of \doubref\EYQ\LS, while in \hsect3,
after summarizing the relevant parts of the \nex2 supersymmetric
free-field realization, we will construct the gravitational
excitations in terms of free-field vertex operators and find a
simple algebraic characterization of the complete spectrum.

\chapter{Equivariant cohomology in
 topological matter-gravity systems}

We will define topological $W_\M$-gravity coupled to
$W_\M$-matter rather formally by thinking of it as a multi-field
generalization of the theories introduced in
\nrf{\topgr\Keke\DVV}\refs{\topgr{--}\DVV}. That is, the Liouville,
matter and ghost sectors are separately topologically twisted \nex2
superconformal systems, each consisting of $2\M\!-\!2$ superfields
\LS. We will denote the corresponding currents of the topological
algebra by $J_{(i)},G_{(i)}^+,G^-_{(i)},T_{(i)}$, where
$i=l,gh,m$\foot{These subscripts refer to Liouville, ghost and matter
sectors respectively.}. There are two anti-commuting \brs\ charges,
$$
\eqalign{
Q_s\ &=\ \cont G_{tot}^+\ \equiv\
\cont(G_{(l)}^++G_{(m)}^++G_{(gh)}^+)\cr
Q_v\ &=\ \cont d\theta^+\Big(\sum_{j=1}^{\M-1}C_{-j}\tilde
W_{j+1}\Big)\ ,
}\eqn\QSVdef
$$
where $C_{-j}$ is the $j$-th ghost superfield with spin $-j$ and
$\tilde W_{j+1}\equiv D_-(W_{j(l)}+W_{j(m)}+\shalf W_{j(gh)})$, and
where $W_{j(i)}$ is the spin $j$ current superfield in the sector
$(i)$. Note that due to the non-linearity of the $W$-algebra,
$W_{j(gh)}$ is very complicated for $j>1$ \doubref\LS\BLLS, and
depends upon both the matter and Liouville fields.

\ni Following \doubref\EYQ\LS\ we introduce the operator
$$
S\ =\ \exp\Big[\cont\Big(\sum_{j=1}^{\M-1}c_{-j}\tilde
U_{j+1}\Big)\Big]\ ,
\eqn\Sdef
$$
where $c_{-j}\equiv C_{-j}|_{\theta=0}$ and $\tilde U_{j+1}\equiv
\tilde W_{j+1}|_{\theta=0}$. One can then show that
$$
\eqalign{
S^{-1}Q_{tot}S\ &\equiv\ S^{-1}(Q_s+Q_v)S\ =\ Q_s\cr
S^{-1}b_{j+1}S\ &=\ b_{j+1}+U_{j+1} \ .\cr
}\eqn\rot
$$
The cohomology problem is thus reduced to the computation of the
cohomology of $Q_s$. Moreover, the condition for equivariant
cohomology: $b_{{j+1},0}\ket{\rm state}=0$, which yields the
operators of topological $W_\M$-gravity, is changed to
$$
(b_{{j+1}}+U_{j+1})_0\ket{\rm state}\
=\ 0\ ,\qquad \ \ j=1,\dots,\M-1.
\eqn\modifequ
$$
At this point it should be stressed that, strictly speaking, the
foregoing has been explicitly verified only for $\M=2$ \EYQ\ and
$\M=3$ \LS. However, we think it is highly plausible that it is true
in general. Thus, we will assume in the following that the physical
spectrum of topological $W_\M$-gravity coupled to $W_\M$-matter can
be obtained for general $\M$ from the cohomology of $Q_s$ modulo the
equivariance conditions \modifequ.

The problem of finding matter representatives of all physical states
is now considerably simplified since $Q_s$, and the equivariance
condition, do not mix the matter, Liouville and ghost Hilbert
spaces. Thus a pure matter physical state, $\ket X$, must satisfy
$$
\eqalign{
Q_{s,(m)}\ket X\ &=\ 0\cr
(U_{j+1,(m)})_0\ket X\ &=\ 0\ ,\ \ \ \ j=1,\dots,\M-1,\cr
}\eqn\physscond
$$
and it is trivial precisely if there is some $\ket Y$ such that:
$$
\eqalign{
\ket X\ =\ Q_{s,(m)}\,&\ket Y\ ,\ \ {\rm with }\cr
(U_{j+1,(m)})_0&\ket Y\ =\ 0\ ,
\ \ \ \ j=1,\dots,\M-1\ .
}\eqn\triv
$$
A further simplification is obtained if we work with states
that are simultaneously highest weight states of the
\nex2 \sca and eigenstates of the zero modes of the super-$W_\M$
currents. Then the equivariance conditions reduce to
$(G^-_{(m)})_0\ket{\rm eigenstate}=0$, since the other conditions
follow from the commutation relations of the super-$W_\M$ algebra.

The physical operators of topological $W$-gravity are the
gravitational descendants, $\ga p$, $p=1,\dots,\M-1$, and polynomials
thereof \topw. They are \brs-exact, $\ga
p=\{Q_s,\del^pc_{-p}\}$, but since $\del^pc_{-p}$ do not
satisfy the equivariance conditions, the $\ga p$ are non-trivial
operators. To find corresponding representatives in the matter
sector, $\Ga p(z)$, we must find operators $\L_p(z)$ such that $\Ga
p(z)=\{Q_s,\L_p(z)\}$ and such that $(\del^pc_{-p}(z)-\L_p(z))$
satisfies the equivariance conditions. That is, we must have
$$
(U_{j+1,(m)})_0\ket{\L_p}\ =\ a_j\d_{jp}\ ,
\eqn\condx
$$
for some non-zero constants $a_j$. In particular, this means that in
terms of the twisted \nex2 superconformal algebra, we have
$$
\Ga1\ =\ G^+_{-1/2}\L_1\ ,\quad\ \ \
{\rm but}\quad \ \ G^-_{1/2}\L_1\ =\ {\rm const}\ .
$$
This directly violates the Hodge decomposition theorem \LVW\ for
unitary \nex2 \sc\ theories, and therefore establishes that a matter
representation of gravitational descendants can only be found in a
non-unitary extension of the matter system.

For the particular representation of these theories (with $\M=2,3$)
in terms of \LG models, it was shown in \doubref\EYQ\LS\ that
arbitrary polynomials in the \LG\ variables provides a natural
extension of the matter Hilbert space, and yields the requisite
matter representation of the gravitational descendants.

\chapter{Free-field realization}

To illustrate our ideas we will consider the \nex2 $W_3$ minimal
models,
which are based on cosets $SU(3)_k\over U(2)$ \KS. They can be
constructed via Drinfeld-Sokolov reduction from $SU(3|2)$ in terms
of free superfields \Wred; in particular, the super-$W_3$ generators
were
explicitly given in \doubref\Wred\DeNiB. Here we will give simply the
(left-moving) \nex2 superconformal generators in component fields:
$$
\eqalign{
J(z)\ &=\ [\p1\bp1+\p2\bp2-i\,
    \a_0(\df1+\df2-\bdf1-2\bdf2)](z)\cr
G^+(z)\ &=\ \sqrt2\big[\bp1\df1+\bp2\df2+
     i\,\a_0(\del\bp1+2\del\bp2)\big](z)\cr
G^-(z)\ &=\ \sqrt2\big[\p1\bdf1+\p2\bdf2+
     i\,\a_0(\del\p1\,+\,\del\p2)\big](z)\cr
T(z)\ &=\ \sum_{j=1}^2\big[-\df j\bdf j +
     \shalf(\p j\del\bp j+\bp j\del\p j)\big](z)\cr
&\ \  -\coeff i2\a_0\big(\del^2\phi_1+\del^2\phi_2+
    \del^2\bar\phi_1+2\del^2\bar\phi_2\big)(z)
}\eqn\eqA
$$
where $\a_0=1/\sqrt{k\!+\!3}$ and where we have used the conventions:
$\phi_i(z)\bar\phi_j(w) \sim -\d_{ij}\log(z-w)$,
$\psi_i(z)\bar\psi_j(w) \sim -\d_{ij}(z-w)^{-1}$. The
Drinfeld-Sokolov reduction also provides the screening operators that
are needed to truncate the free-field spectrum. These screening
operators fall into two classes, the $D$-type screeners,
$$
\eqalign{
R_1^D\ &=\ \big[(\df1-\df2+\bdf2)+2i\,\a_0(\p1-\p2)\bp2\big]
   e^{-i\a_0(\phi_2-\phi_1+\bar\phi_2)}\cr
R_2^D\ &=\ \big[(\df1-\bdf1)+2i\,\a_0\p1\bp1\big]
   e^{-i\a_0(\phi_1+\bar\phi_1)}\cr
S_1^D\ &=\ \big[(\df1-\df2+\bdf1)-2i\,\a_0(\p1-\p2)\bp1\big]
   e^{i\a_0(\phi_2-\phi_1+\bar\phi_1)}\ ,\cr
}\eqn\eqB
$$
and the $F$-type, or fermionic, screeners:
$$
\eqalign{
P_1^F\ &=\ i\p1e^{{i\over{a_0}}\phi_1}\ ; \qquad\qquad\qquad\ \ \ \
P_3^F\ =\ i\bp1e^{{i\over{a_0}}\bar\phi_1}\cr
P_2^F\ &=\ i(\p2-\p1)e^{{i\over{a_0}}(\phi_2-\phi_1)}\  ;\ \ \ \ \ \
P_4^F\ =\ i\bp2e^{{i\over{a_0}}\bar\phi_2}\cr
}\eqn\eqC
$$
The unitary, minimal superconformal model is obtained by using a
\brs\
operator constructed from all of the $D$-type screeners, and half of
the $F$-type screeners, for example $P_1^F$ plus $P_2^F$.
The combined topological matter-gravity systems can thus be
described by combining this free-field \brs\ operator with
the \brs\ operator $Q_s+Q_v$ described in \hsect2.

The idea is now to perform the \brs\ reduction in a slightly
different manner. Namely, first perform the $D$-type \brs\ reduction,
and combine the $F$-type screening charges obtained from $P_1^F$ and
$P_2^F$ with the supercharge $Q_s$ in \QSVdef. One can perform the
$F$-type \brs\ reduction at any point, because the \brs\ charge is
local with respect to the other screening currents, and anti-commutes
with $Q_s$ and the $D$-type screeners. (Note that one can view the
left-moving $F$-type screeners as the left-moving part of the
``right-moving'' supercharge $\tilde G^+(z)$ (and similarly for
$G^+(z)$), such that they behave exactly like the terms
$\sum\psi_i{\del W\over\del x_i}$ of the supercurrents in the \LG
formulation \LGfree.)

Before we discuss the reduction method in more detail, we think it
valuable to say more precisely what is the underlying idea of
performing the $D$-type reduction while not performing the $F$-type
reduction. A simple way to understand this is to consider the \nex2
superconformal minimal models, whose coset form is $SU(2)_k\times
SO(2)\over U(1)$. If one uses the Kac-Wakimoto realization of
$SU(2)_k$, one introduces a superghost system, $(\b,\g)$, plus a
single free boson. To form the coset model, one bosonizes the
super-ghosts according to \FMS
$$
\eqalign{
\b\ &=\ (\del\xi)e^{-\phi}\ =\ i(\del\chi)e^{i\chi-\phi}\cr
\g\ &=\ \ \ \eta e^\phi\ \ \ =\ e^{-i\chi+\phi}
}\eqn\fmsbos
$$
To recover the Hilbert space of the $(\b,\g)$ system one has to fix
the momenta $p_\phi-p_\chi$ and exclude all states involving the zero
mode, $\xi_0$. Equivalently, one can allow states with
$p_\phi-p_\chi\geq0$, and then compute the cohomology of the
fermionic charge $Q=\oint\eta$ \BMP. This is precisely what is
accomplished by the fermionic screener $P^F$ in the superconformal
model. That is, if one does not employ the fermionic screener, one
obtains infinitely many copies of the physical ``matter'' Hilbert
space, related by spectral flow in $p_\phi-p_\chi$. From the
following it will be clear that these copies can be reinterpreted as
gravitational descendants of the matter sector.

\ni According to what was said above, we take
$$
Q_{s,(m)}\ =\ \cont G^+(z) - i\cont(P_1^F+P_2^F)(z)\ .
\eqn\Qsm
$$
One easily finds that
$$
\eqalign{
e^{{i\over\a_0}\phi_1}\ &=\ \big\{Q_{s,(m)},(\bp1+\bp2)\big\}\cr
e^{{i\over\a_0}(\phi_2-\phi_1)}\ &=\ \big\{Q_{s,(m)},\bp2\big\}\ .\cr
}\eqn\star
$$
Moreover, one has
$$
G^-_{1/2}\p j= \ i\sqrt2\,\a_0\ =\ {\rm const.}
$$
This makes the vertex operators in \star\ into candidates for
matter representatives of the pure $W$-gravity descendants, and
according to our considerations in \hsect2,
we simply need to verify equation \condx. This is most simply
accomplished by showing that there are linear combinations
of the $\p i$ that are eigenstates (with distinct eigenvalues)
of the zero mode, $S_0$, of the lowest, spin-2 component of the
$W_3$ supercurrent.

The easiest way to isolate this spin-2 current is to remember that
the \nex2 super coset model factorizes according to:
$$
{SU(3)_k\times SO(4)\over SU(2)_{k+1}\times U(1)}\ \cong
{SU(3)_k\over SU(2)_k\times U(1)}\otimes
{SU(2)_k\times SU(2)_1\over SU(2)_{k+1}}\otimes U(1)\ .
\eqn\factor
$$
The second factor is a Virasoro minimal model, and its Coulomb gas
realization is embedded in the \nex2 free-field formulation
above \doubref\DeNiA\DeNiB.
Specifically, define a scalar field, $\vp$, by
$$
\del\vp\ =\ -\sqrt{\coeff{k+3}{2(k+2)}}(\del H_1-\del H_2)-
\coeff1{\sqrt{2(k+2)}}(\df1-\df2-\bdf1)\ ,
\eqn\vpdef
$$
where $\p j=e^{i H_j}$, $\bp j=-e^{-i H_j}$. The stress tensor
$T(z)$ in \eqA\ contains a part that is precisely
$$
T^\vp\ =\ -\shalf(\del\vp)^2 -
\coeff i{\sqrt{2(k+2)(k+3)}}\del^2\!\vp\ ,
$$
and this is the natural candidate for the stress tensor
of the minimal model. There is, however, a subtlety. The operator
$T^\vp$ does not quite commute with the screening charges
of the \nex2 theory. The operator that does commute
with the screening charges rather is:
$$
\eqalign{
\hat T^\vp(z)\ &=\ T^\vp(z) + N(z)\ ,\ \ \  {\rm where}\cr
N(z)\ &=\ \coeff i{(k+2)\sqrt{k+3}}(\p1\bp2)\bdf1 -
 \coeff1{k+2}(\del\p1)\bp2\ .}
\eqn\hatTN
$$
The origin and interpretation of this extra term will be
described in \DeNiB. For the present we simply observe that
it modifies the computation of the eigenstates.

\ni In terms of the decomposition \factor, the spin-2 current
is then given by
$$
S(z)\ =\ T(z) - \coeff1{4k(k+3)}J^2(z) -
 \coeff{(k+2)(5k-3)}{k(k+5)}\hat T^\vp(z)\ .
\eqn\Sdef
$$
This is a good conformal field, with normalization $S(z)S(0)\sim
{(5k-3)(2k+3)(k-1)\over k(k+3)(k+5)}\coeff1{z^4}$. From \Sdef\ we
find that $\bp2$ and $\bp1+\bp2$ are eigenstates of $S_0$, with
eigenvalues: $-{(k^2+3)\over k(k+5)}$ and $-1$, respectively. This
proves that indeed certain linear combinations of the matter vertex
operators \star\ are \brs\ representatives of the gravitational
excitations, $\ga1,\ga2$.

In the \nex2 \sc model, the elements of the primary chiral ring
(which constitutes the ``matter'' sector) can be very simply
represented by\foot{Note that these expressions are much simpler than
the cumbersome expressions for the ground ring elements given in
\BLNWB, where a different free-field realization was used.}
$$
V_{a,b}\ =\ e^{{i\over\sqrt{k+3}}[a\phi_1+b(\phi_2-\phi_1)]}\ ,
\eqn\vertex
$$
for $0\leq b\leq a\leq k$. Based on the fact that $\ga1$ and $\ga2$
can be represented by $V_{k+3,0}$ and $V_{0,k+3}$, we expect that the
gravitational descendants can be represented by the vertex operators
$V_{a,b}$ with $a,b \geq 0$.  We will show that this is the case, but
that one needs to make the further restriction: $0 \leq b \leq a$ in
order to get one copy of each of the gravitationally dressed physical
states. (These ``gravitational'' ring elements will be chiral, but
not primary conformal fields.)  The problem is to determine which
of these vertex operators correspond to physical states and which
correspond to null states. There are two parts to this. First, we
must consider the equivariant cohomology. Obviously:
$$
\eqalign{
V_{a,b}\ &=\ \big\{Q_{s,(m)},(\bp1+\bp2)\,V_{a-(k+3),b}\big\}\cr
         &=\ \big\{Q_{s,(m)}, \bp2\,V_{a,b-(k+3)}\big\}\ ,\cr
}\eqn\VRexpand
$$
and one can verify that
$\L^{(1)}_{a',b'}\equiv(\bp1+\bp2)V_{a',b'}$ and
$\L^{(2)}_{a',b'}\equiv\bp2V_{a',b'}$ are eigenstates of $S_0$.
It therefore suffices to check whether $G^-_{1/2}
\L^{(i)}_{a',b'}$ are physical states or not. If either
$\L^{(1)}_{a-(k+3),b}$ or $\L^{(2)}_{a,b-(k+3)}$ is null,
then so is $V_{a,b}$. This allows us to determine all physical
states in a recursive way, exactly as in \LS.

One easily sees that $G^-_{1/2} \L^{(2)}_{a,b-(k+3)}\equiv0$ if
$b\!=\!k\!+\!2$ and $a$ is arbitrary. Similarly, $G^-_{1/2}
\L^{(1)}_{a-(k+3),b}\equiv0$ if $a\!=\!k\!+\!1$ and $b$ is
arbitrary. These lines of unphysical states are repeated for $a$
and $b$ modulo $(k\!+\!3)$, via the recurrence property just
mentioned. The general scheme is displayed in \lfig\figone.
\figinsert\figone{Lines of null states cut through the domain
$0\leq b\leq a$ of physical vertex operators, $V_{a,b}$. We indicated
the {\it raison d'\^etre} for each of the lines. Note that the lines
are repeated modulo $(k\!+\!3)$. The bottom triangle represents the
primary chiral ring of the matter model, and the other triangles
represent the gravitational excitations.}{1.5in}{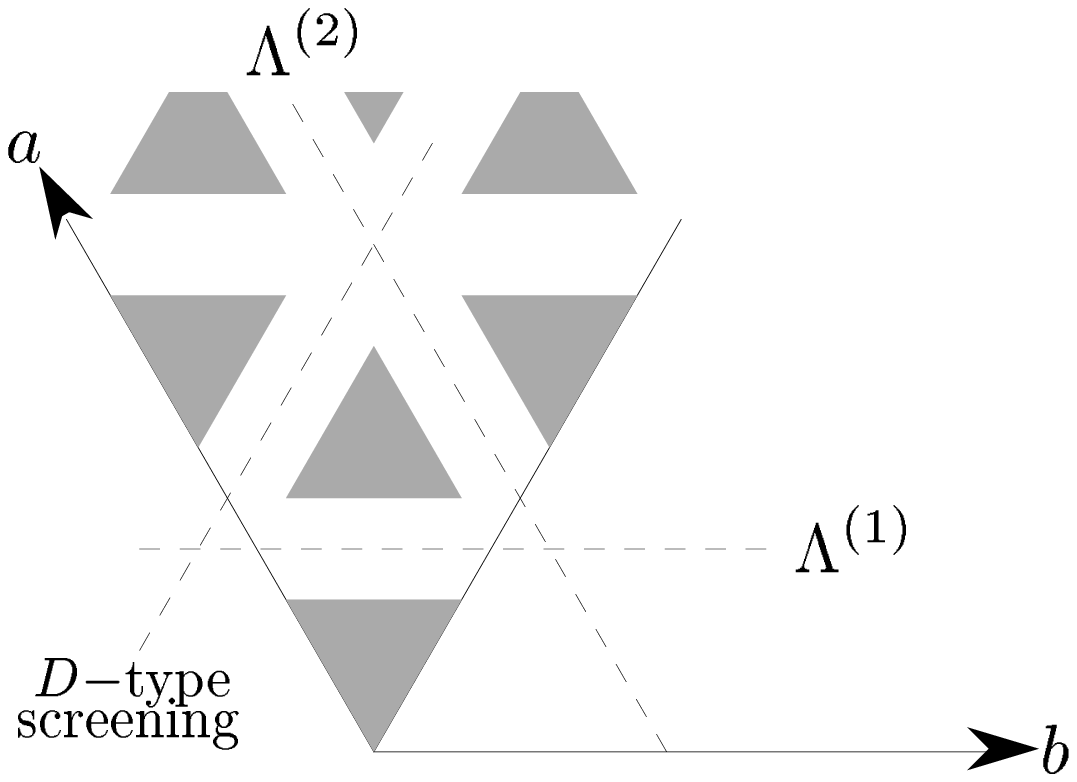}

There is one further line of unphysical states (also repeated modulo
$(k\!+\!3)$). This line appears because of the $D$-type screening in
the \nex2 \sc models. To see this, remember that the complete set of
screeners, \eqB\ plus \eqC, is obtained via Drinfeld-Sokolov
reduction from the roots of the super-algebra $SU(3|2)$. In
particular, $R^D_1$ and $R^D_2$ are associated with the bosonic
$SU(3)$ sub-algebra, while $S_1^D$ is associated with the $SU(2)$
sub-algebra. Indeed, a detailed analysis
\refs{\DeNiA,\DeNiB} shows that $S_1^D$
contains the \brs\ current exp$({i\sqrt{{2(k+2)\over k+3}}\vp})$
corresponding to the $SU(2)$ denominator factor in the
minimal model part of the decomposition \factor. Thus, this $D$-type
screening constructs the minimal model Hilbert space using the
denominator screening operator. There is, however, a subtlety,
related to the presence of the nilpotent part $N(z)$ of $\hat
T^\vp(z)$ in \hatTN. The fact that the quantum numbers of the of the
numerator factor of ${SU(2)_k\times SU(2)_1\over SU(2)_{k+1}}$ in
\factor\ are correlated with the quantum numbers of ${SU(3)_k\over
SU(2)_{k+1}\times U(1)}$ means that the model has a special modular
invariant. In particular, the total Hilbert space of the model can be
mapped back onto itself by essentially trading of minimal model
$\vp$-momentum translations against translations in the momenta of
the Coulomb gas realization of ${SU(3)_k\over SU(2)_{k+1}\times
U(1)}$. These $\vp$-momentum translations are associated with the
numerator factor of the minimal model ${SU(2)_k\times SU(2)_1\over
SU(2)_{k+1}}$. The fermionic screening is then precisely what further
reduces the Hilbert space to a single copy of the physical states.
These issues will be discussed more fully in \DeNiB.

The practical consequences of the foregoing in the present context is
that because we have not performed the $F$-type screening, we must
allow all values of the $\vp$-momentum. However, because we have
performed \brs\ reduction using the denominator screening charge as
part of the $D$-type reduction, we have to take into account the
corresponding null states. For general vertex operators, $e^{ip\vp}$,
this has two effects. First, one restricts to $p\geq0$, or
equivalently, one keeps only one state for each eigenstate of the
zero mode of $\hat T^\vp(z)$. Secondly, for $p\!=\!\coeff
n{\sqrt{2(k\!+\!2)(k\!+\!3)}}$, one gets null states for
$n\!=\!(k\!+\!2)[1\!+\!2m(k\!+\!3)]$, $m\in\ZZ$. In terms of the
\nex2 \sc fields,  one sees from \vpdef\ that $\phi_2$ is
orthogonal to $\vp$, while
$\del\vp(z)\df1(w)\sim - \coeff1{\sqrt{2(k+2)}}\coeff1{(z-w)^2}$,
hence
$V_{a,b}$ has $\vp$-momentum equal to
$p\!=\!\coeff{a-b}{\sqrt{2(k+2)(k+3)}}$. The restriction $p\geq0$
means that we must take $a\geq b$, and the line of null states occurs
at $a\!-\!b\!=\!(k\!+\!2)$. This means that the corresponding
$V_{a,b}$ are unphysical.

With \lfig\figone\ we have thus recreated the picture of physical
states found in \LS. However, our present discussion provides a more
direct link to the structure of the underlying Lie algebra than the
formulation in terms of \LG fields. Specifically, we can associate to
$V_{a,b}$ a weight of $SU(3)$ via:
$$
\l_{a,b}\ =\ b\,\l_1 + (a-b)\, \l_2 + \rho\ ,
$$
where $\l_1,\, \l_2$ are the fundamental weights of $SU(3)$, and
$\rho=\l_1\!+\!\l_2$ is the Weyl vector. Let now $\a_1$ and $\a_2$ be
a
system of simple roots. The requirement that $\l_{a,b}$ lie in the
interior of the the fundamental Weyl cone is:
$$
\l_{a,b}\cdot\a_j\ >\ 0\ ,\ \ \ j=1,2\ ,
\ \ \ \ {\rm or}\ \ \ a\geq b\geq0\ .
\eqn\fundcha
$$
If we also require that
$$
\l_{a,b}\cdot\a\ \not\equiv\ 0\ \mod\,(k+3)
\eqn\lines
$$
for any root, $\alpha$, of $SU(3)$, we arrive at
$$
\big\{\,a-b\ \not\equiv k+2\ ,\
b\ \not\equiv k+2\ ,\
a\ \not\equiv k+1\,\big\}\,\mod(k+3)\ ,
$$
which are precisely the physical state conditions that we evolved
above.

The conditions \fundcha\ and \lines\ are indeed well-known in the
representation theory of affine algebras, and just determine the
highest weights of $\hat{SU}(3)$ (which are a subset of the highest
weights of $SU(3)$). Note that the states of the topological matter
model (those with $0\leq b\leq a\leq k$) correspond to the subset of
integrable highest weights, whereas their gravitational descendants
correspond to non-integrable highest weights, \ie, to non-unitary
representations. This is in accordance with our general
considerations at the end of \hsect2.

The generalization of the foregoing to arbitrary \nex2 $W_\M$ models
is quite obvious, though it might be difficult to compute the details
explicitly. The underlying Drinfeld-Sokolov reduction of
$SU(\M|\M\!-\!1)$ (with canonical gradation) leads to $(2\M\!-\!1)$
$D$-type and to $(2\M)$ $F$-type screeners, and the corresponding
coset decomposes as
$$
{SU(\M)_k\times SO(2n-2)
\over SU(\M\!-\!1)_{k+1}\!\times\! U(1)}\cong
{SU(\M)_k\over SU(\M\!-\!1)_{k}\!\times\! U(1)}\otimes
{SU(\M\!-\!1)_k\!\times\! SU(\M\!-\!1)_1\over
 SU(\M\!-\!1)_{k+1}}\otimes U(1)\ .
\eqn\factor
$$
Performing the $D$-type screening but not the $F$-type screening will
give infinitely many copies of the ``matter'' Hilbert space involving
translations of the momenta in the $(\M\!-\!1)$ directions of the
$SU(\M)$ weight lattice. The $D$-type screening, along with
positivity of the $U(1)$ charge, will however restrict the momenta in
the positive dual Weyl cone of $SU(\M)$, and will introduce lines of
null states associated with $SU(\M\!-\!1)$. The other null states can
then be identified by imposing equivariance. The result will almost
certainly be that the physical states of topological $W_\M$-matter
coupled to topological $W_\M$-gravity are generated by vertex
operators associated with the ($\rho$-shifted) $\hat{SU}(\M)$ highest
weights, $\l$, which satisfy
$$
\l\cdot\a\ >\ 0\ ,\ \ \ \ \l\cdot\a\ \not\equiv\ 0\ \mod (k+\M)
$$
for all positive roots, $\a$, of $SU(\M)$. We thus arrive at a very
simple and satisfying algebraic characterization of the physical
states, or ground rings, in these theories.
\goodbreak
\refout
\end